# Empirical Study of Phased Model of Software Change

Leon A. Wilson, Yoann Senin, Yibin Wang, Václav Rajlich
Wayne State University
Detroit MI 48202, USA
{leon.wilson3, yoann.senin, yibin.wang, rajlich}@wayne.edu

*Abstract*—Software change is the basic task of software evolution and maintenance. Phased Model for Software Change (PMSC) is a process model for software changes that localize in the code. It consists of several phases that cover both program comprehension and code modifications. This paper presents an empirical study of an enactment of PMSC, enhanced by the use of tool JRipples. The subjects are graduate students with varying degree of programming experience. The empirical findings demonstrate that programmers with knowledge of PMSC and supported by JRipples perform perfective software changes in unfamiliar software in significantly less time (about half time) than unaided programmers. Substantial time improvements were witnessed in both code comprehension and implementation efforts.

## I. INTRODUCTION

Software evolution and maintenance consist of repeated software changes that consume a substantial amount of programmer effort [1, 2]. Although there has been considerable research conducted in code changes, most of it has been focused on subtasks and less attention has been given to the integrated software change process itself. The research of the integrated software change process can contribute to better understanding of software evolution and improve the productivity and quality of software development. An adoption of software process models in other contexts has demonstrably led to such improvements [3].

In this article, we report an exploratory empirical study that evaluates effectiveness of programmers using the Phased Model for Software Change (PMSC) assisted by a specialized tool and compares their performance against unassisted programmers who are not following any predefined software change process. PMSC is geared towards software changes that localize in the code and their purpose is to introduce new functionality into software, while preserving or increasing the quality of the code.

The rest of this paper is organized in the following manner: Section 2 surveys prior work. Section 3 provides an example of enactment of PMSC process. Section 4 presents empirical study of programmers following PMSC in contrast to unassisted programmers. The results are discussed in Section 5. Finally, conclusions and future work are reported in Section 6.

## II. PREVIOUS WORK

A preliminary process model of software change appeared in [4]. A newer software change process model is the test-driven development (TDD) where the developers implement the new functionality in two phases: in the first phase, they write tests for the new functionality, and in the second phase, they write code that passes these tests [5]. Feathers discusses software change in a complex legacy code and deals with many special situations [6].

PMSC subsumes these earlier models. An example of actual software change, i.e. the enactment of PMSC, appeared in [7]. It was further investigated in the context of Solo Iterative Process [8, 9] and education [10, 11]. A detailed explanation of PMSC appeared in [12]. The role of PMSC in software evolution and maintenance is emphasized in a survey paper [13], where the acronym "PMSC" was used for the first time. This paper provides further insight and assesses the impact of PMSC on developer productivity. In order to make the paper self-contained, we briefly explain PMSC.

PMSC consists of phases or subtasks summarized in Figure 1. Note that some of these subtasks have received a large attention of the researchers, and the state of the research has been summarized in surveys; interested readers are directed to these surveys. More detail is also available in Chapters 5 – 11 of [12].

First phase of PMSC is *initialization* and it deals with requirements, their elicitation, analysis, prioritization, and so forth. The outcome of this phase is a selection of a particular change request or bug report for implementation. These issues are discussed in [14, 15].

During *concept (or feature) location* phase, programmers locate the code units where the new or corrected functionality will appear [16, 17]. These code units implement concepts that are relevant to the change request or defect report.

PMSC continues by *impact analysis* that identifies all source code units affected by the proposed change [18-20]. It allows programmers to assess the overall impact and difficulty of the change and to choose an appropriate implementation strategy.

*Actualization* implements the new functionality and incorporates it in the appropriate place that was found during concept location. Change propagation is also conducted during actualization to ensure that any secondary change is properly done throughout the program. During the conclusion phase, the revised source code is committed to the software repository.

*Prefactoring* reorganizes the existing code in order to make the change easier. *Postfactoring* eliminates technical debt introduced into the program during actualization. Both of these phase are special instances of refactoring [21, 22].

During all phases where code modifications occur, software *verification* certifies the quality and correctness of the code [23, 24].

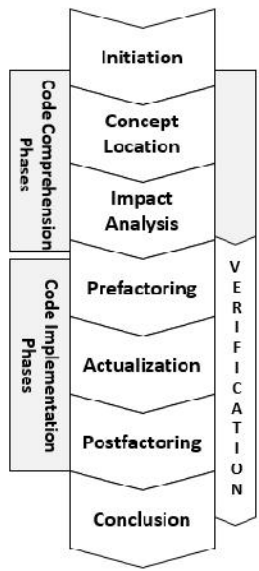

Figure 1. Phases of PMSC

The tool *JRipples* is an Eclipse plug-in that supports parts of PMSC, namely concept location, impact analysis, and change propagation (part of actualization). It is available from http://jripples.sourceforge.net/.

### III. EXAMPLE OF PMSC ENACTMENT

PMSC enactment contains some or all phases. As an example of PMSC enactment, consider a simple program that supports Point of Sale in a small store, see the UML class diagram in Figure 2.

(1) Initialization

The selected change request is "The system supports cash-only payments. Add a support for credit card payments."

(2) Concept Location

The programmer extracts the concept "payment" from the change request. After using a suitable concept location technique, the programmer finds that the concept is located in the Payment class.

(3) Impact Analysis

Class Payment interacts with the Sale class. The programmer inspects code of Sale and concludes that it is impacted by the change. Then the programmer explores the neighbors of the class Sale that consists of classes Session, Store and SaleLineItem. These classes are not impacted by the change, so impact is limited to classes Payment and Sale.

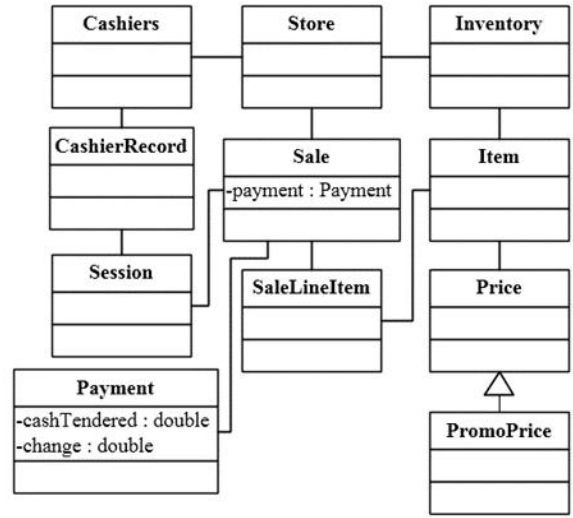

Figure 2. Point of Sale (PoS) System

(4) Prefactoring

The class Payment in Figure 2 supports only cash payment. In order to make the change easier, the programmer decides to extract from this class a new superclass AbstractPayment that contains both data and algorithms that are shared by all types of payments.

(5) Actualization

In actualization, the programmer adds a new class Credit that is derived from AbstractPayment, and also makes secondary changes in class Sale.

(6) Postfactoring

The classes now have illogical names. To rectify that, class Payment is renamed Cash and class AbstractPayment is renamed Payment.

(7) Conclusion

The code after change undergoes system verification and then new versions of all changed files are committed to software repository.

Figure 3 is the final result of this software change. Corresponding changes are also made to the testing harness.

Figure 4 shows a JRipples window during impact analysis in this example. After the programmer marks the classes Payment and Sale as "Impacted", JRipples automatically marks classes Session, Store and SaleLineItem by a mark "Next". These classes interact with impacted classes and should be inspected, in order to determine whether they also may be impacted.

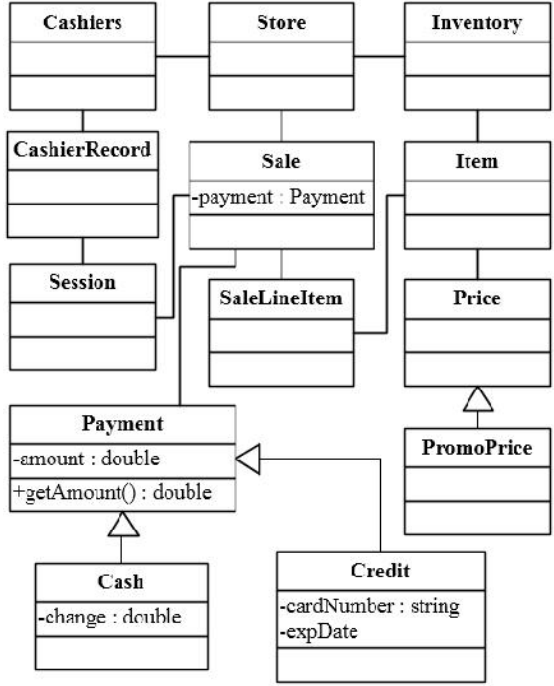

Figure 3. PoS after the change

Figure 4. JRipples in impact analysis

## IV. EMPIRICAL STUDY

To investigate the effectiveness of PMSC supported by JRipples, we conducted an empirical study. The control was an unaided software change where developers used only the standard Eclipse and their own self-taught process; there was no instructions what process they should use. The treatment was PMSC instruction and the use of JRipples. The study design followed generally accepted practices in software engineering empirical research, found in [25]. We posit the following alternative hypotheses:

H1: PMSC + JRipples shorten the completion time of software changes.

H2: PMSC + JRipples shorten the code comprehension portion of the time of software changes.

H3: PMSC + JRipples shorten the code implementation portion of the time of software changes.

### A. Study Design

We conducted an earlier pilot study [26]. The experience from that study helped us to design the experiment reported in this paper. We performed within-subject experiment [27] comparing the performance of unassisted programmers against the same programmers aided by PMSC and JRipples. That is, we performed a two-stage experiment, using the first stage of the experiment as a baseline performance. Then we compared that with the performance under the treatment. As reported in [27], a within-subject design provides a higher degree of experimental control since study participants serve as their own control group.

Stage 1 was the control stage, and Group 1 participants were assigned change request tasks from System A, while Group 2 were assigned change request tasks from System B. Stage 2 was the treatment stage, and participants switched systems. The same set of change requests were used within both stages. Change requests from each system were randomly assigned. This helped to reduce any potential biases within the change request assignments. Between the stages, we trained the participants in use of PMSC and JRipples.

During the entire study, two assistants gave limited supports to participants by answering questions, providing clarity, as well as monitoring/evaluating participants' work. The assistants screened each participant's work and interviewed the students in order to certify the quality of the data.

### B. Subjects of Study

There were 17 participants. They included both M.S and Ph.D. students in their first semester of their graduate studies. All participants were required to complete a pre-study questionnaire in order to ascertain their programming experience and the familiarity with systems and supporting tools used in the study.

All participants had an earlier experience with programming and with software changes. All were familiar with Java and all had experience with at least one other programming language. Their previous programming experience ranged from 0.75 to 4 years. They had no previous knowledge of the source code of subject applications. They also did not know PMSC or JRipples before taking part in the study. They were randomly assigned to one of the two groups.

After the first stage, programmers were provided with an orientation on the PMSC process and the JRipples tool. For the training, we used Chapters 6 – 10 of [12]. We also assigned a homework that provided a hands-on experience in both PMSC process and JRipples.

Finally, participants performed a post-study survey. It included rating the difficulty of performing the changes, the opinion whether following PMSC and using JRipples saved them time, and any additional general comments. Assistants again were in place to guarantee the validity and quality of all reported data and work.

*C. Objects of Study*

We used two open source software applications:

jEdit is a Java-based text editor available from http://www.jedit.org. The version of jEdit used for this study includes approximately 100 kLOCs, 850 classes and 517 files.

JabRef is a Java-based application allowing one to store and manage journal references and is available from http://jabref.sourceforge.net/. The version used for this study includes approximately 78 kLOCs, 835 classes and 577 files.

Note that the applications are from different domains in order to reduce any bias caused by the knowledge the subjects obtained in the previous stage. Table 1 lists all the change requests used in this study.

TABLE 1 CHANGE REQUESTS

| | JEdit |
|---|---|
| 1 | Under menu Search, add menu item "mark all". Locate all matches and add markers to all of the lines. |
| 2 | Currently there are no timestamps in the activity log file. Add timestamps to all kinds of messages. |
| 3 | Currently jEdit shows a red dot at the end of every line. Newline is the only whitespace symbol that jEdit shows. Add menu item Show/Hide whitespace under menu View to allow the user to choose whether all whitespace symbols (newlines, blanks, and tabs) will be shown. |
| 4 | Currently jEdit allows users to access the text that was previously searched by pressing page up or right-click keys in Search Dialog. Display in a listbox the last 5 text fragments that were previously searched. |
| 5 | Allow the user to specify a signature to be used as the footer in all printed documents. An option should be available to enable/disable the signature. When the option is enabled, the signature will appear in the status bar. |
| | **JabRef** |
| 1 | Input: a folder, output: a .bib file<br>Create a GUI for this functionality. |
| 2 | Input: a .bib file, a folder containing .tex files<br>Output: a new .bib file<br>Create a GUI for this functionality. |
| 3 | The current format of the timestamp when adding an entry is [year].[month].[day] (e.g. 2013.11.18). Make a change so that the timestamp has the format [year][month][day].[hh][mm][ss] (e.g. 20131118.083025) and it is auto updated when the button auto is clicked. |
| 4 | User can click on the table column to sort. Save the order to the bibFile.<br>Input: A BibTeX file<br>Output: A BibTex file, with items sorted as viewed with Jabref table |
| 5 | In the feature "Autogenerate BibTeX keys" keys are generated in this format [author][year]. Make a change so that the BibTeX keys have the timestamp added to the format like this [author][year]_[hhmmss] |

*D. Supporting Tools*

The Eclipse version 4.2.2 was the IDE. Subversion and TortoiseSVN were selected as the version control system and end user client to manage the overall changes to all applications among multiple programmers. Participants used Rabbit (an Eclipse plug-in) to track the amount of time spent performing various activities within the Eclipse IDE. The programmers used their own laptops and worked on their tasks during their own time.

Upper time limits were set at 16 hours for completing code comprehension and 16 hours for code implementation.

V. EMPIRICAL RESULTS

Table 2 presents data from the study. The columns contain for Stage 1 and Stage 2 contain time spent in minutes. In order to select the appropriate statistical test, we analyzed whether the data is normally distributed. To do so, we calculated the time effort differences (i.e. total amount of time of control group minus total amount of time with treatment), and performed an Anderson-Darling normality test on the resulting data. The normality test produced $p = 0.27$ greater than $\alpha = 0.05$, indicating normal distribution. Therefore, we used the Paired T-test parametric test with a 95% confidence interval. Note that the Paired T-test is commonly used when measuring "Before" and "After" results if using the same participants.

TABLE 2 USER STUDY DATA

| Participant | | | Stage 1 | | | Stage 2 | | |
|---|---|---|---|---|---|---|---|---|
| ID | Yrs Exp | Yrs Java | Code Compre | Code Imple | Total | Code Compre | Code Imple | Total |
| **Group 1** | | | | | | | | |
| 1 | 2 | 0.5 | 195 | 86 | 281 | 125 | 291 | 416 |
| 2 | 4 | 1 | 720 | 120 | 840 | 135 | 30 | 165 |
| 3 | 2 | 0.5 | 480 | 135 | 615 | 45 | 20 | 65 |
| 4 | 0.75 | 0.5 | 60 | 120 | 180 | 290 | 497 | 787 |
| 5 | 1 | 0.5 | 300 | 180 | 480 | 360 | 130 | 490 |
| 6 | 2.5 | 0.5 | 300 | 360 | 660 | 53 | 70 | 123 |
| 7 | 0.75 | 0.5 | 480 | 480 | 960 | 55 | 90 | 145 |
| 8 | 2 | 1 | 300 | 120 | 420 | 160 | 285 | 445 |
| 9 | 2 | 1.5 | 20 | 20 | 40 | 110 | 85 | 195 |
| **Group 2** | | | | | | | | |
| 10 | 3 | 1.5 | 180 | 780 | 960 | 90 | 360 | 450 |
| 11 | 1.75 | 0.75 | 960 | 960 | 1920 | 90 | 25 | 115 |
| 12 | 1 | 0.5 | 30 | 75 | 105 | 100 | 35 | 135 |
| 13 | 2 | 0.5 | 90 | 150 | 240 | 105 | 130 | 235 |
| 14 | 2 | 1 | 120 | 360 | 480 | 45 | 85 | 130 |
| 15 | 1.5 | 0.25 | 480 | 15 | 495 | 70 | 94 | 164 |
| 16 | 1.5 | 0.5 | 938 | 427 | 1365 | 49 | 95 | 144 |
| 17 | 1.5 | 0.5 | 700 | 380 | 1080 | 280 | 200 | 480 |
| Mean for All | | | 373.71 | 280.47 | 654.18 | 127.18 | 148.35 | 275.53 |
| St. Dev for All | | | 305.65 | 266.81 | 491.64 | 94.57 | 134.78 | 198.39 |

## A. PMSC Performance Results

To establish and confirm the above alternative hypotheses, we formulated and tested the corresponding null hypotheses. For hypothesis H1, data in Table 2 show that software changes took an unassisted programmer (during Stage 1) on average 654.18 minutes, while following PMSC and using JRipples (during Stage 2) they took 275.53 minutes. This constitutes 57.88% ($p = 0.033$; $t = 2.31$) improvement in total time. This is a statistically significant result and we attest hypothesis H1.

For hypothesis H2, data in Table 2 show that code comprehension took an unassisted programmer on average 373.71 minutes, while following PMSC took 127.18 minutes. This constitutes a 65.97% ($p = 0.012$; $t = 2.78$) improvement in time required for program comprehension. This is a statistically significant result and we attest hypothesis H2.

For hypothesis H3, data in Table 2 show that the implementation of software changes took an unassisted programmer on average 280.47 minutes, while following PMSC took 148.35 minutes. This constitutes a 47.11% ($p = 0.218$; $t = 1.28$) improvement in the time for implementation. This test does not provide conclusive statistical evidence to reject the null hypothesis. It should be noted that although we cannot substantiate statistically hypothesis H3, the average reduction in implementation time is large.

When analyzing programmers' logs, we noticed that in stage 1, participants spent large amount of time trying to comprehend code that was unrelated to the software change. In stage 2, their comprehension effort was much better targeted due to the use of concept location and impact analysis of PMSC and use of JRipples tool. That may partially explain the large improvement in code comprehension between the stages.

We also noticed two outliers. We reviewed the log of participant 11 and conducted a follow-up interview. The subject had a lot of difficulties in Stage 1 determining where to make the coding change (i.e. code comprehension) and spent significant time inspecting GUI objects, source code unitts and so forth. We further investigated the performance of other participants who were assigned the same change request and found that others completed that change request in a more reasonable time; hence the problem was not in assignment of a particularly difficult change. In Stage 2, participant 11 performed better than other participants who were assigned the same task, i.e. instruction in PMSC and use of JRipples turned out to be particularly beneficial to this subject.

The other outlier was subject 9 who in Stage 1 performed noticeably faster than other participants assigned the same change request. After reviewing the relevant logs and conducting follow-up interviews, we found that the change request was not overly simple but the participant was very effective due to a lucky insight. It should be expected that programmer performance on any given change request will vary. However, note that in stage 2 the performance of all programmers was more uniform.

## B. Additional Observations

In post-study survey, results in Table 3 show on second row that 13 out of 17 programmers (76%) self-reported that performing software changes following PMSC was more effective and saved time.

TABLE 3 PMSC POST STUDY DATA (PART 1)

| Question | Yes | No |
|---|---|---|
| Was PMSC Effective? | 15 | 2 |
| Did PMSC Save Time? | 13 | 4 |

We also found that programmers on average reported that both Stage 1 and Stage 2 change requests are of about the same difficulty (see Table 4).

TABLE 4 PMSC POST STUDY DATA (PART 2)

| On a scale 1 – 5, rate the difficulty of software change | Avg. | ±† |
|---|---|---|
| Stage 1 | 3 | 0.93 |
| Stage 2 | 3 | 1.07 |

The research in program comprehension is motivated by large proportion of the developer's time that is devoted to comprehension. However, the hard data on that proportion are scarce. For that reason, we also computed the ratios of the time of code comprehension to the total time for all 34 changes in Table 2, and summarized them in the boxplot in Figure 5. Please note that the median value of the ratio is 50% and it roughly corresponds to the frequently quoted Corbi's estimate [28] and that there is a wide variation between individual changes.

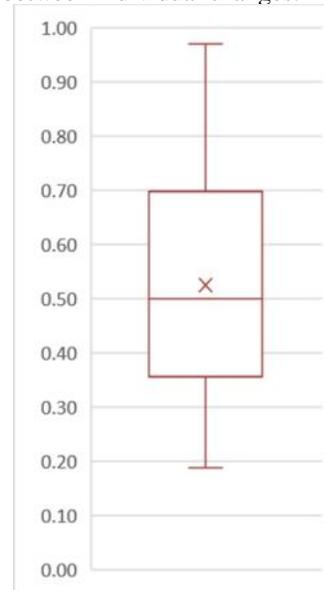

Figure 5. Program comprehension share of the software change effort

*C. Threats to Validity*

Several issues can affect the interpretation of the results. The confounding factors include the participant's years of programming experience, the familiarity with the object applications, complexity of the requested changes, learning during the experiment, or the supporting tools. In order to partially mitigate these potential confounding factors, we used the same set of change requests in both stages and assigned them randomly. We also ensured that the supporting tools were consistent between the stages as well as between the participants groups. We also partially mitigated learning effect by using two very dissimilar systems within the experiment design and using programmers with previous experience, for whom software change was not a new experience.

As for construct validity, we ensured that the quantitative data were properly collected, and also collected pre-experiment and post-experiment qualitative data. Admittedly, the self-reporting of performance has limitations and to mitigate that, we reviewed the logs from the participants. We required that the participants demonstrate that their software changes are running correctly; by that, we made sure that the work was completed and had the expected quality.

Also note that as a treatment, we used the PMSC process supported by tool JRipples and hence we did not measure separate contribution of the process and the tool. The questions regarding the separate contribution of the tool and the process have not been answered by our experiment.

Regarding external validity, specific programming technologies, application domains, subject programs and change requests may impact the results. Also the performance improvement among programmers of varying experience may differ. In particular, programmers who are familiar with the application may spend considerably less time in the comprehension part. In order to lessen the impact of that, we made sure that our programmers were unfamiliar with both applications.

In our experiment, we used students as subjects. The current literature provides conflicting evidence about the validity of such studies. In certain software engineering circumstances, the use of students in lieu of professional programmers produced only minor difference in performance [29]. However in other circumstances, a distinct difference exists between novice and expert programmers, such as dealing with exception handling [30] . We submit that our graduate students with previous professional programming (rather than inexperienced undergraduate students) are closer to the expert programmers. However their range of experience is only 0.5 – 1.5 years and hence extension of these results to experienced software developers may present another external validity problem.

## VI. CONCLUSION AND FUTURE WORK

The purpose of this research was to investigate the impact of a well-defined phased model for software change (PMSC) and its supporting tool JRipples on developer productivity. We observed graduate students who implemented a set of software changes in software that they were unfamiliar with, using their own personal unassisted processes (control) versus using the PMSC and JRipples (treatment). We found statistically significant evidence that PMSC with JRipples substantially shortened the time taken to complete a software change (by 58%), particularly in the phases of program comprehension (by 66%). Thus we found strong evidence that PMSC supported by JRipples is an effective process. We also replicated the evidence that program comprehension consumes approximately half of the software change effort.

In the future, we plan to broaden the experiment to the subjects from industry. We also plan to conduct a similar study with subjects that are familiar with the software that will be changed and see the impact of PMSC + JRipples in this context. We also want to conduct experiments that would separate our treatment into two, i.e. impact of unsupported PMSC and impact of the tool JRipples.

Another future research topic is to evolve JRipples in order to seamlessly cover additional tasks of PMSC, particularly refactoring and testing. Such more powerful tool may further improve the productivity gains of programmers.